\documentclass[twocolumn]{article}

 \pdfoutput=1

\usepackage{graphicx}
\usepackage{amsmath}
\usepackage{authblk}

\title{Topological complexity of photons' paths in biological tissues}

\author[1,2,*]{Tiziano Binzoni}
\author[3]{Fabrizio Martelli}
\author[4]{David Cimasoni}

\affil[1]{Department of Basic Neurosciences, University of Geneva, Geneva, Switzerland}
\affil[2]{Department of Radiology and Medical Informatics, University Hospital, Geneva, Switzerland}
\affil[3]{Dipartimento di Fisica e Astronomia dell'Universit\`a degli Studi di Firenze, Sesto Fiorentino, Firenze, Italy}
\affil[4]{Universit\'e de Gen\`eve, Section de math\'ematiques, 2 rue du Li\`evre, Gen\`eve, Switzerland}
\affil[*]{Corresponding author: tiziano.binzoni@unige.ch}

\begin{document}

\maketitle

\begin{abstract}
In the present contribution three means of measuring the geometrical and topological complexity of photons' paths in random media are proposed.
This is realized by investigating the behavior of the average crossing number, the mean writhe, and the minimal crossing number
of photons' paths generated by Monte Carlo (MC) simulations, for different sets of optical parameters.
It is observed that the complexity of the photons' paths increases for increasing light source/detector spacing,
and that highly ``knotted'' paths are formed.
Due to the particular rules utilized to generate the MC photons' paths, the present results may have an interest
not only for the biomedical optics community, but also from a pure mathematical point of view.
\end{abstract}

\section{Introduction}
\label{sec:Introduction}

In many research and clinical fields, light is used to non-invasively extract 
information on what is under the surface of the investigated medium \cite{ref:Durduran2010,ref:Katz2011,ref:Wiersma2013}.
Thus, knowing the paths followed by photons inside the medium is extremely important.
In fact, photons can carry information only from regions of the medium that they have ``visited''.
For this reason, the study of, e.g., the penetration depth of photons migrating in random media 
represents a fundamental topic 
\cite{ref:Bonner1987,ref:Nossal1988,ref:Weiss1989,ref:Cui1991,ref:Feng1995,ref:Patterson1995,ref:Wiess1998,
ref:Weiss1998b,ref:Weiss1998c,ref:Bicout1998,ref:Bicout1998a,ref:Gandjbakhche2000,ref:Carp2004,ref:Zonios2014,ref:Martelli2016}.

Photons' penetration depth may be limited for two reasons, that have different consequences on the measurements:
1) the photons' path is short; 2) the photons' path is long but folds many times like a "ball of wool".
From a practical point of view, even if photons reach exactly the same depth, 
photons in case 2 visit a larger region of the medium, due to the length of their path.
This means that, when detected, photons do not necessarily carry the same information in both cases.

For paths of equal lengths, different geometrical characteristics of photons' paths 
may also be sufficient to influence the detected signals.
This may be the case in diffuse  correlation spectroscopy \cite{ref:BoasCampbell1995}
or in laser-Doppler flowmetry \cite{ref:BonnerNossal1981},
where more or less ``contorted'' paths influence in different manner the indirectly detected ``phase'' changes of laser light.

From the theoretical side, knowing the typical geometrical characteristics of the photons' paths
may represent an added value.
In fact, several techniques allowing to modelize photons transport in random media,
such as the path integral approach \cite{ref:Jacques1998,ref:Polishchuk1997}, assume some implicit
hypothesis on the geometry and more or less ``smooth'' behavior of the most probable paths.
Having a clear understanding of the actual paths shapes may therefore help to improve these models. 

These are few reasons describing why it would be useful to be able to better 
understand the geometrical characteristics of photons' path in random media.
To the best of our knowledge, no criteria have been proposed allowing to 
define the ``degree of complexity'' of the photons' paths.
Moreover, it is not clear which percentage of the detected photons may represent photon's path 
with a given degree of complexity.

To this aim, we propose in the present work three ways to measure the geometrical and topological complexity of the photon's paths. These methods are based on three tools borrowed from the mathematical study of knotted curves known as {\it knot theory} \cite{ref:Rolfsen1976,ref:Adams1994,ref:Kawauchi1996,ref:Lickorish1997}:
1) the average crossing number ($\bar c$); 2) the mean writhe ($\bar w_r$) ; and 3) the minimal crossing number.
The proposed analysis has obviously not the pretension to be an exhaustive treatment of the problem,
but should be seen as a first contribution in this direction.

\begin{figure}[t]
\includegraphics[width=\linewidth]{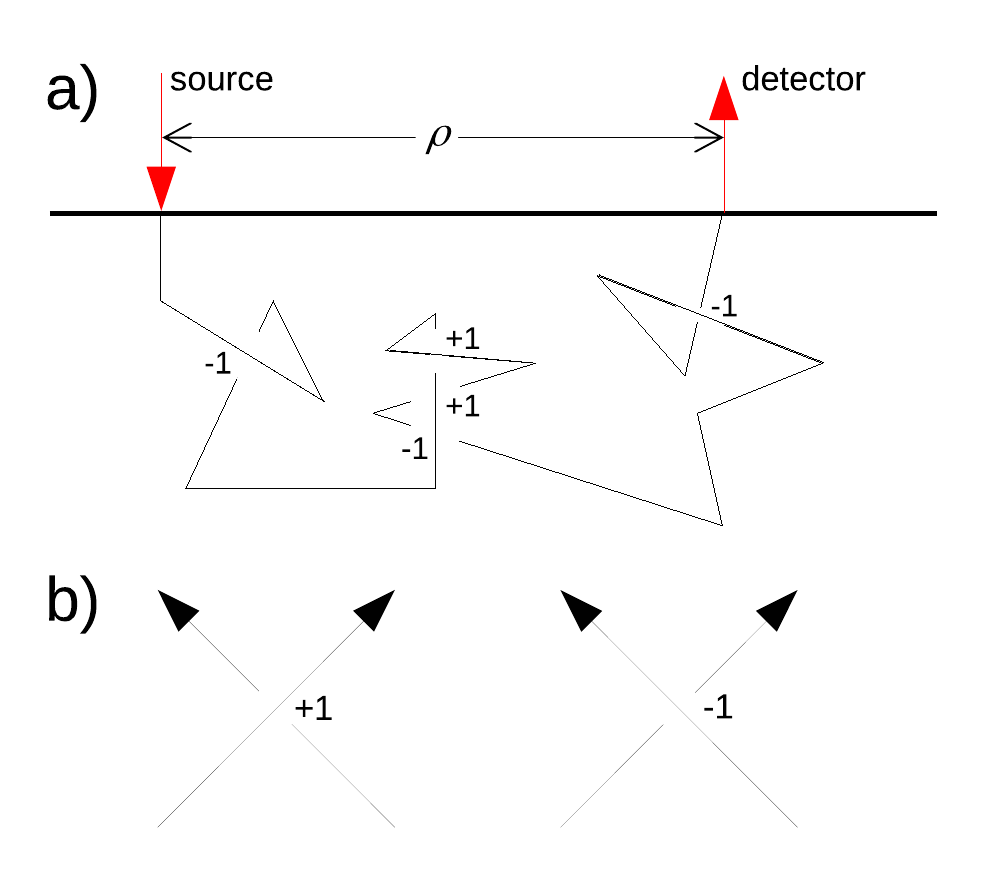}
\caption{a) A photon's path from a certain vantage point. At each crossing, the broken line should be understood as passing ``under'' the continuous line. The orientation of the photon's path goes from the source to the detector, defining the direction of the arrows on each crossing. 
b) Positive and negative crossings, defined using the orientation of the path.}
\label{fig:figurePath}
\end{figure}

The {\it average crossing number} of a photon's path can be informally defined as follows. If an observer looks at the path from a fixed position, he will see a curve (Fig. \ref{fig:figurePath}a)
with a certain number of crossings, which of course depends on the position of the observer. Averaging this number over all the vantage points gives the average crossing number $\bar c$ of the path
(more technically, it is obtained by integrating the crossing number over the set of directions, i.e. the~$2$-dimensional 
unit
sphere, and dividing the result by~$4\pi$, the area of the sphere).
It is intuitively obvious that $\bar c$  measures the geometrical complexity of the path: higher the $\bar c$ value, more complex the entanglement of the "ball of wool".

If the path under study is oriented, and a photon's path is, then one can define its {\it mean writhe} in an manner analogous to the average crossing number. Simply observe that the orientation of the path allows to associate a sign ($+1$ or~$-1$) to each observed crossing (see Fig.~\ref{fig:figurePath}b): the sum of these signs is called the writhe, and averaging this number over all the vantage points defines the mean writh~$\bar w_r$ of the path.
In this case, the geometrical meaning of this number is more subtle. In a nutshell,~$\bar w_r$ tells us if the path turns more like a right ($\bar w_r>0$) or left ($\bar w_r<0$) handed helix. Note that a path can be very much entangled and still have vanishing mean writhe: this only means that it behaves equally as right and left handed helices.  

Finally, the {\em minimal crossing number} of a given closed path can be defined as the minimal number of crossings, from all vantage points, of any closed curve obtained by continuously deforming the given one. The physical intuition behind this mathematical notion is the following: the closed path, or {\em knot}, should be though of as a piece of rope with both ends spliced together, and continuously deforming it means trying to disentangle the rope without cutting it. The minimal crossing number tells us how close we can get to completely disentangling the rope; in other words, it measures the topological complexity of the knot.
 
Determining the minimal crossing number of an arbitrary knot is very difficult, and can be seen as one of the founding questions in the mathematical study of knots. Fortunately, this theory provides us with a variety of tools, some of which can be used to evaluate the minimal crossing number (see Sec.~\ref{subsec:HOMFLY}). 

Strictly speaking, a photon's trajectory is a path from the source to the detector, and therefore not a closed loop. However, there is a canonical way to close this path: simply connect the detector back to the source with a line segment (see Fig~\ref{fig:figurePath}a). In this way, each photon's path yields a closed loop, i.e. a knot, whose topological complexity can be studied via its minimal crossing number. 

\section{Methods}
\label{sec:Methods}

The photons' paths have been generated for a semi-infinite geometry (Fig. \ref{fig:figurePath}a) and a pencil beam 
light source impinging normally on the boundary of the medium.
The point detector was situated at different distances $\rho$ from the source.
The refractive index external to the medium (e.g. air) was set to 1.
The absorption coefficient ($\mu_{\rm a}$) and the reduced scattering coefficient ($\mu_{\rm s}'$) 
were set to 0.025 mm$^{-1}$ and 1 mm$^{-1}$, respectively.
These values may represent a typical biological tissue such as the human skeletal muscle \cite{ref:Torricelli2004}.

\subsection{Photons' paths generation}
\label{subsec:PhotonsPaths}

The photons' paths were generated by Monte Carlo (MC) simulation using the 
well known MCML approach \cite{ref:Wang1995,ref:SassaroliMartelli2012}.
The algorithm was implemented in Matlab\textsuperscript{\textregistered} language.
The paths were stored in a file, together with their weights, 
by saving the generated sequence of the 3D-coordinates of the scattering points
(points of reflections on the boundaries included).
A Henyhey-Greenstein scattering phase function was utilized \cite{ref:HenyeyGreenstein1941}.
MC simulations were generated for four different sets of optical parameters, i.e.
refractive index of the medium ($n$) and anisotropy factor ($g$), and for seven $\rho$ values
(see Sec. \ref{sec:Results} for the explicit values).
Around 19\,000 paths were generated for each MC simulation.

\subsection{Mean writhe}
\label{subsec:Writhe}

The mean writhe $\bar w_r$ of a general (piecewise smooth) curve is difficult to compute exactly.
However, photons' paths are polygonal curves, 
and a simple and exact formula exists for this specific case \cite{ref:Levitt1983,ref:Klenin2000}.
The algorithm was implemented in Matlab\textsuperscript{\textregistered}
language (see Sec. ``Method 1a'' in Ref. \cite{ref:Klenin2000}
for the equations) and tested on so-called ``ideal knots'', where $\bar w_r$ is known~\cite{ref:Stasiak1998}.
The database for these ideal knots was taken from~\cite{ref:ScottDror}.
A $\bar w_r$ value was obtained for each MC simulated  photon's path.

\subsection{Average crossing number}
\label{subsec:acn}

For each MC simulated  photon's path, the average crossing number was obtained 
by means of the same Matlab\textsuperscript{\textregistered} code utilized to assess $\bar w_r$: one simply needs to take the absolute value of the summed terms~$+1$ and~$-1$ in the programming code to obtain $\bar c$ instead of~$\bar w_r$ \cite{ref:RogenFain2003}.

\subsection{Minimal crossing number and HOMFLY polynomial}
\label{subsec:HOMFLY}

As explained in the introduction, each photon's path defines a knot, whose minimal crossing number can be very difficult to compute in general. We now elaborate on this topic by reviewing some knot theory, before explaining our strategy to overcome these difficulties.

\begin{figure}[htbp]
\includegraphics[width=\linewidth]{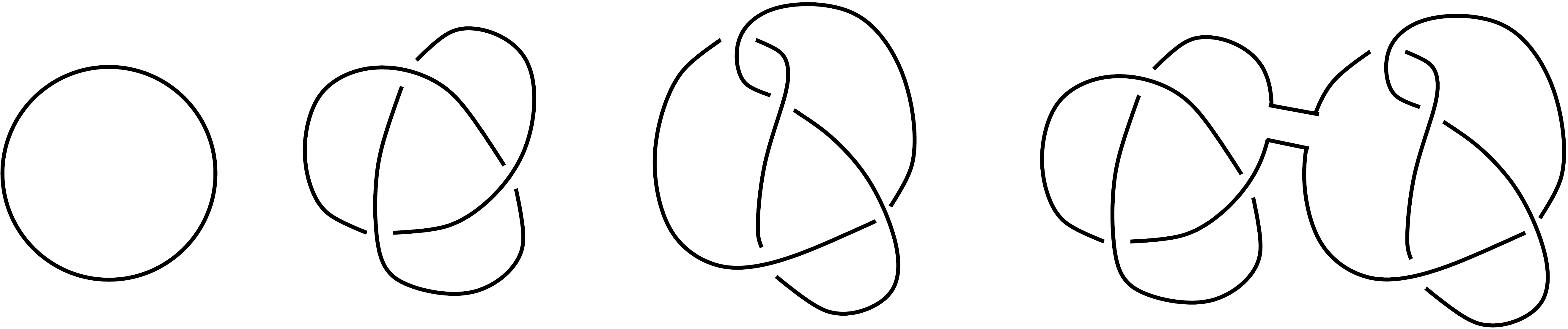}
\caption{Four examples of knots: the unknot~$O$, the trefoil knot~$3_1$, the figure-eight knot~$4_1$, and the connected sum of a trefoil and a figure eight knot, denoted by~$3_1\#4_1$.}
\label{fig:figureKnots}
\end{figure}

First observe that a knot~$K$ has minimal crossing number~$0$ if and only if it can be disentangled to give the {\em unknot}~$O$ illustrated in Fig.~\ref{fig:figureKnots}.
In such a case, one says that~$K$ and~$O$ are {\em equivalent}, or that they define the same {\em knot type}. There is no knot with minimal crossing number~$1$ or~$2$, and a unique (up to mirror image) knot type with minimal crossing number~$3$, called the trefoil knot (see~Fig.~\ref{fig:figureKnots}). Proving that this knot~$K$ indeed had minimal crossing number equal to~$3$, and not less, amounts to showing that it is not equivalent to the unknot, i.e. that~$K$ is a {\em non-trivial knot}. Even though this claim is intuitively clear, as trying to disentangle such a piece of rope quickly shows, rigorously proving it is no easy task: it requires the help of a quantity associated to each knot that remains unchanged under continuous deformations. If such a {\it knot invariant} takes different values for~$K$ and for~$O$, this proves that~$K$ cannot be deformed to the unknot.

Fortunately, knot theory provides us with many such invariants. For obvious reasons, the ''value'' of an invariant is judged upon two criteria: its computability, and its capacity of telling apart many knots. If it existed, a ''perfect invariant'' would be easy to compute, and {\it complete}, i.e. take different values for any two non-equivalent knots. One specific invariant is excellent in this respect, although not perfect, and this is the so-called HOMFLY polynomial~\cite{ref:HOMFLY1985,ref:PT1988}. We shall not present formally this polynomial, but mention that its very definition gives an algorithm for its computation (see however~\cite{ref:Jaeger1990} for a proof that a fast algorithm is very unlikely to exist). Let us also note that, even though it is not a complete invariant, the HOMFLY polynomial is extremely powerful: in particular, it is widely believed to ``detect the unknot'', i.e. to take the specific value~$1$ only for the unknot.

We therefore computed the HOMFLY polynomial of each photon's path, writing an R code exploiting the well tested R-package (``Rkots'') presented in Refs.~\cite{ref:ComoglioRinaldi2012,ref:ComoglioRinaldi2011}.
From this data, it is very easy to tell the number of non-trivial knots: simply count the number of HOMFLY polynomials different from~$1$. Furthermore, the number of different HOMFLY polynomials gives a lower bound on the number of different knot types, and this bound is very close to the actual number. 

What about the (minimal) crossing number? Several tables are available that list knots up to~$15$ crossings together with the values of many invariants, including the HOMFLY polynomial (see e.g.~\cite{ref:ScottDror} and references therein).
Following the convention of Alexander and Briggs~\cite{ref:AlexanderBriggs1926}, most tables list and denote the knot types according to their crossing number: first comes the trefoil knot, denoted by~$3_1$, then a unique 4-crossing knot, known as the figure-eight, and denoted by~$4_1$ (see Fig.~\ref{fig:figureKnots}). There are~$2$ different knot types with crossing number equal to~$5$, denoted by~$5_1$ and~$5_2$, then~$3$,~$7$,~$21$,~$49$ and~$165$ distinct knot types  with crossing number~$6$,~$7$,~$8$,~$9$ and~$10$, respectively. Overall, there are exactly~$313\,230$ knot types with at most~$15$ crossings. Therefore, given the HOMFLY polynomial of a photon's path, it is possible to look for the first knot type in this table with corresponding HOMFLY polynomial, and hence determine its crossing number.

There is an issue with this strategy: the tables do not list all knot types, but only {\it prime knots} -- and the numbers given above are actually the numbers of prime knot types. These are knots that cannot be expressed as the {\it connected sum}~$K_1\#K_2$ of two non-trivial knots~$K_1$ and~$K_2$, an operation illustrated in Fig.~\ref{fig:figureKnots}. Fortunately, 
any knot can be expressed in a unique way as the connected sum of a finite number of prime knots, and this factorization is unique up to the order of the prime factors. Also, the HOMFLY polynomial 
of~$K_1\#K_2$ is equal to the product of the HOMFLY polynomials of~$K_1$ and of~$K_2$. Finally, the crossing number is widely believed to be additive under connected sum. For example, the following polynomial was computed by analysing a photon path generated by an MC simulation:
\begin{equation*}
P(a,z) = 2 + z^2 - 3a^2 - 3a^2z^2 + 2a^4z^2 + 3a^4 - a^2z^4 - a^6\,.
\end{equation*}
This polynomial factors as the product  of~$-a^4+a^2z^2+2a^2$ and~$a^2-a^{-2}-z^2-1$,
which are the HOMFLY polynomials of the trefoil and figure-eight knots, respectively. Therefore, we obtain the (non-prime) knot~$3_1$\#$4_1$, illustrated in Fig.~\ref{fig:figureKnots}, with crossing number~$3+4=7$.

The amended strategy was then the following, implemented by using a symbolic programming language (symbolic toolbox in Matlab\textsuperscript{\textregistered}). Based on~\cite{ref:ScottDror}, the following three tables were build:
1) A table for HOMFLY polynomials of prime knots with at most~$15$ crossings (unfortunately, the table~\cite{ref:ScottDror} misses 81\,319 of the 253\,293 prime knots with exactly~$15$ crossings, and therefore, so does our table);
2) A table for knots with one, two, or three prime factors each with at most~$10$ crossings; and 
3) A table for knots with exactly four prime factors each with at most~$8$ crossings.
Given a polynomial, we searched these tables for the first knot with corresponding HOMFLY polynomial. Knots that were not found in these tables are not belonging to the three categories listed above: this amounts to a very small proportion of all trajectories, ranging from 0 for small~$\rho$  values to~$\sim 0.1\%$ for~$\rho=30$.

As already mentioned, the HOMFLY polynomial is not a complete invariant: it could happen that the photon's knot type is actually given by another knot, further down the tables, with the same HOMFLY polynomial but higher crossing number. This can happen, but is very unlikely, for three reasons. First of all, the higher its crossing number, the less probable it is for a knot to be generated. Secondly, we are dealing with an extremely powerful invariant. To give an example, among the~$249$ prime knots with at most~$10$ crossings, only four pairs of distinct knots share the same HOMFLY polynomial, namely~($5_1,10_{132})$, ~$(8_8,10_{129})$,~$(8_{16},10_{156})$ and~$(10_{25}, 10_{56})$~\cite{ref:Jones1987}. Finally, there is a wide class of knots, called {\it alternating knots}, for which the crossing number can be deduced directly from the HOMFLY polynomial (see details below). For these knots, it could still happen that we do not get the right knot type, but the crossing number will always be correct. For example, in the pairs listed above, the knots~$10_{129},10_{132}$ and~$10_{156}$ share the same HOMFLY polynomial as a knot with smaller crossing number, and are therefore non-alternating. But this is a relatively rare phenomenon for low crossing number: out of the~$249$ prime knots with at most~$10$ crossings,~$196$ are alternating.

As mentioned above, a huge majority of the photons paths gave HOMFLY polynomials which were found in the above tables.
On the other hand, the proportion of values of HOMFLY polynomials not belonging to these tables is as high as~$40.57\%$. In other words, among the number of different knot types generated, this percentage is formed by knots so intricate that they do not appear in the three tables described above.
There is a way to get some information on the crossing number of these knots as well. Indeed, given the HOMFLY polynomial~$P(a,z)$ of a knot~$K$, one can consider the associated one-variable polynomial~$V(t)=P( t^{-1},t^{1/2}-t^{-1/2})$ known as the {\it Jones polynomial}, and compute its {\it breadth}, i.e. the difference between its highest and lowest degrees. It turns 
out
that this number is a lower bound on the crossing number of~$K$, equal to this number if~$K$ is alternating (see e.g.~\cite{ref:Lickorish1997}). 
Therefore, we also computed the breadth of the Jones polynomial of all the photons' paths, in the hope of finding very complex knots.

\section{Results}
\label{sec:Results}

\subsection{Average crossing number}
\label{subsec:Racn}

\begin{figure}[htbp]
\includegraphics[width=\linewidth]{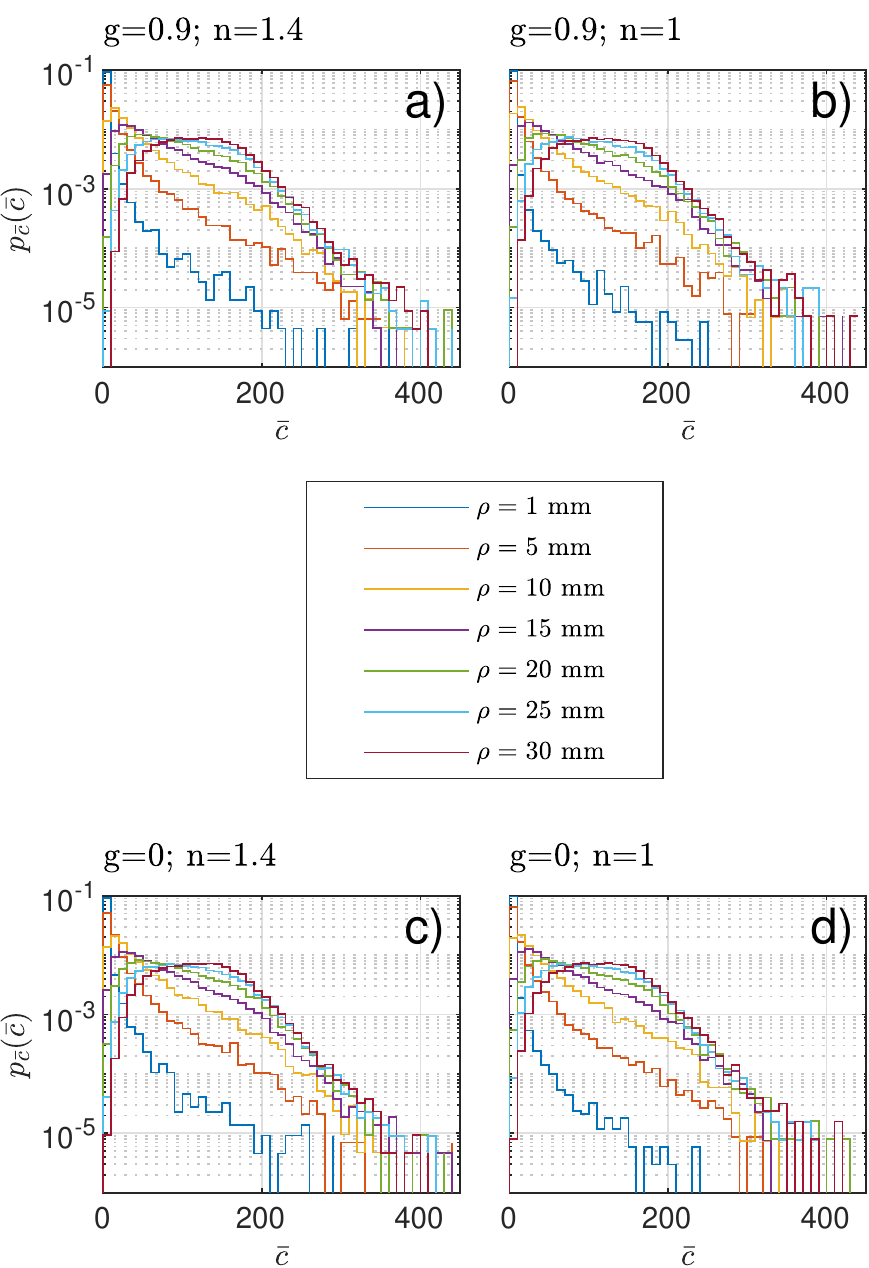}
\caption{Probability density $p_{\bar c}(\bar c)$ that a photon's path
has an average number of crossings close to $\bar c$.
The vertical axis range in the different panels is the same.}
\label{fig:acn}
\end{figure}

In Fig. \ref{fig:acn} 
is reported the probability density function (pdf), $p_{\bar c}$, that a photon's path
has an average crossing number close to $\bar c$.
The $p_{\bar c}(\bar c)$ curves were estimated for four different sets of optical parameters 
(Figs. \ref{fig:acn}a, \ref{fig:acn}b, \ref{fig:acn}c and \ref{fig:acn}d).
It clearly appears that for small $\rho$ values, high $\bar c$ values tend to disappear.
For large $\rho$, the paths become more complex with a high number of crossings 
(see e.g. $\rho=30$ mm).
By comparing Figs. \ref{fig:acn}a, \ref{fig:acn}b, \ref{fig:acn}c and \ref{fig:acn}d it appears that 
$p_{\bar c}(\bar c)$  is not very sensitive to changes in $g$ and $n$.
Thus, for a given $\rho$, the overall ``complexity'' of the paths tends to remain the same
independently of presence of reflections ($n$ value) on the medium boundary, or of the more
or less forward picked phase function ($g$ value).

\begin{figure}[htbp]
\includegraphics[width=\linewidth]{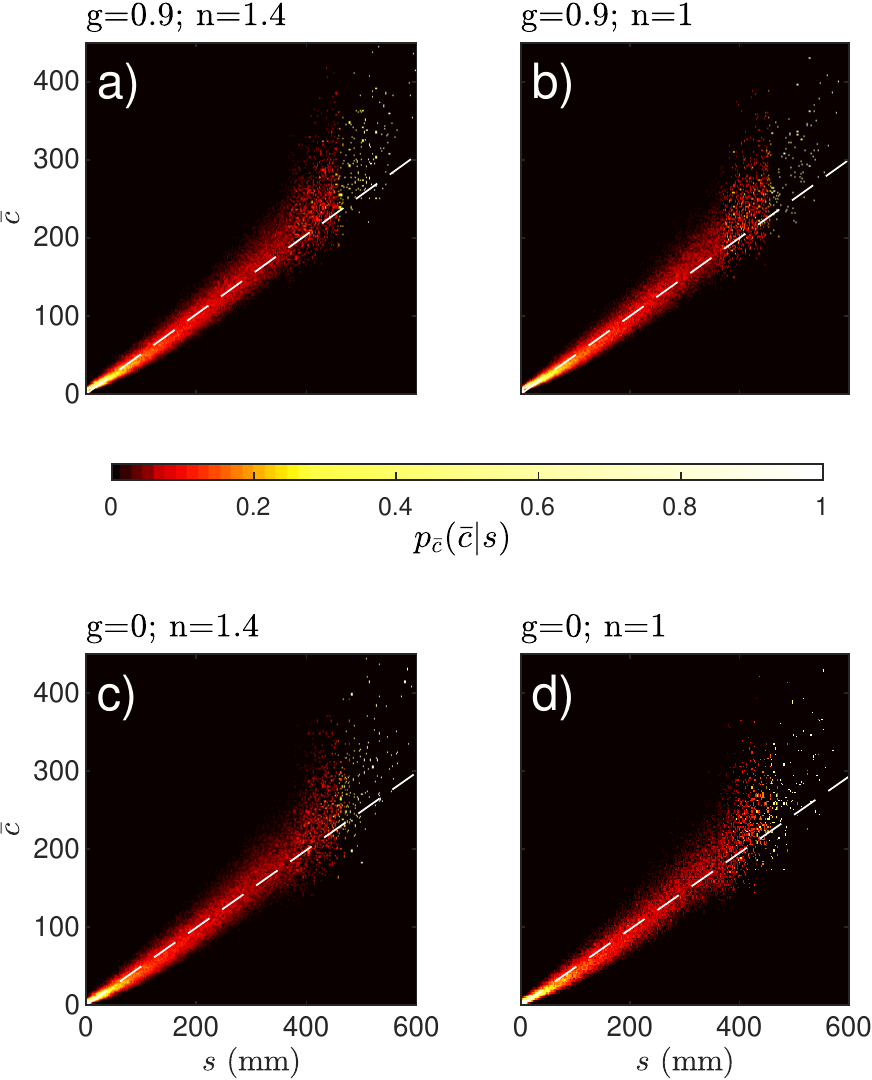}
\caption{Conditional probability density $p_{\bar c}(\bar c|s)$ 
that a photon's path of length $s$ has average crossing number close to $\bar c$
(no distinction is made in this case between different $\rho$ values).
The dashed line is the regression line starting from $(s,\bar c)=(0,0)$, with slopes: 
a) $0.5101 \in [0.5096, 0.5105]$; b) $0.5000 \in [0.4995, 0.5006]$; 
c) $0.4973 \in [0.4968, 0.4979]$ and; d) $0.4885 \in [0.4879, 0.4890]$.
The squared parenthesis define the 95\% confidence intervals.
The horizontal (vertical) axis range in the different panels is the same.}
\label{fig:acns}
\end{figure}

In Fig. \ref{fig:acns}
is reported the conditional pdf, $p_{\bar c}(\bar c|s)$, that a given photon's path of length $s$ has 
$\bar c$ crossings. 
The data in Fig. \ref{fig:acns} are the same as the data in Fig. \ref{fig:acn},
with the distinction between the different $\rho$ values replaced by the distinction between different~$s$ values.
From Fig. \ref{fig:acns}, one can observe that $s$ and $\bar c$ are linearly related.

\subsection{Mean writhe}
\label{subsec:RWrithe}

\begin{figure}[htbp]
\includegraphics[width=\linewidth]{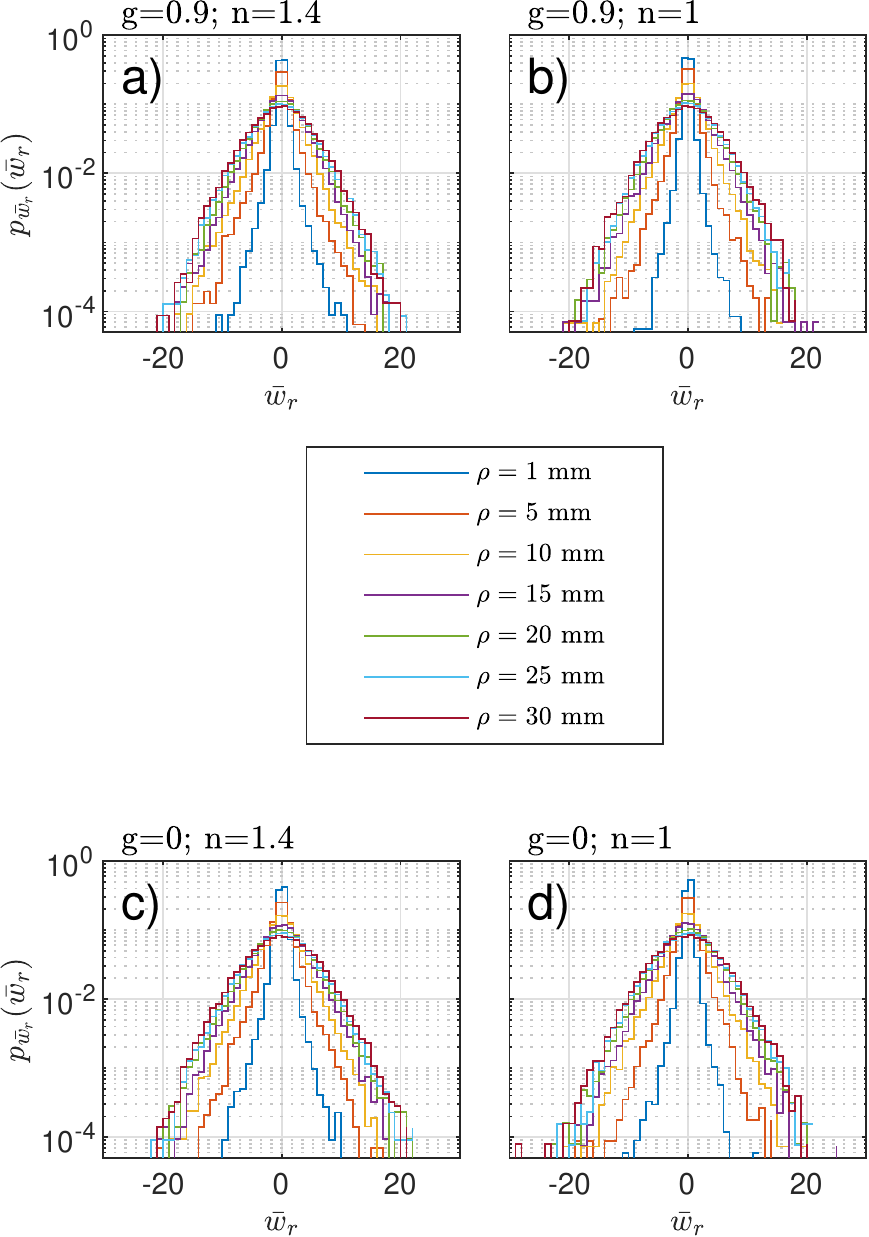}
\caption{Probability density $p_{\bar w_r}(\bar w_r)$ that a photon's path
has a mean writhe equal to $\bar w_r$.
The vertical axis range in the different panels is the same.}
\label{fig:wr}
\end{figure}

In Fig. \ref{fig:wr}
is reported the probability density function, $p_{\bar w_r}(\bar w_r)$, that a photon's path
has a mean writhe close to $\bar w_r$.
The $p_{\bar w_r}(\bar w_r)$ curves were estimated for four different sets of optical parameters 
(Figs. \ref{fig:wr}a, \ref{fig:wr}b, \ref{fig:wr}c and \ref{fig:wr}d).
These curves are clearly symmetric around $\bar w_r=0$, meaning that the paths have the same probability to spiral right or left before reaching the detector.
This is an expected behavior due to the symmetry of the problem.
One can also observe that the larger the interoptode spacing $\rho$ is, the wider the $p_{\bar w_r}(\bar w_r)$ curve is. Furthermore, this behavior does not depend on the optical parameters $g$ and $n$.

\begin{figure}[t]
\includegraphics[width=\linewidth]{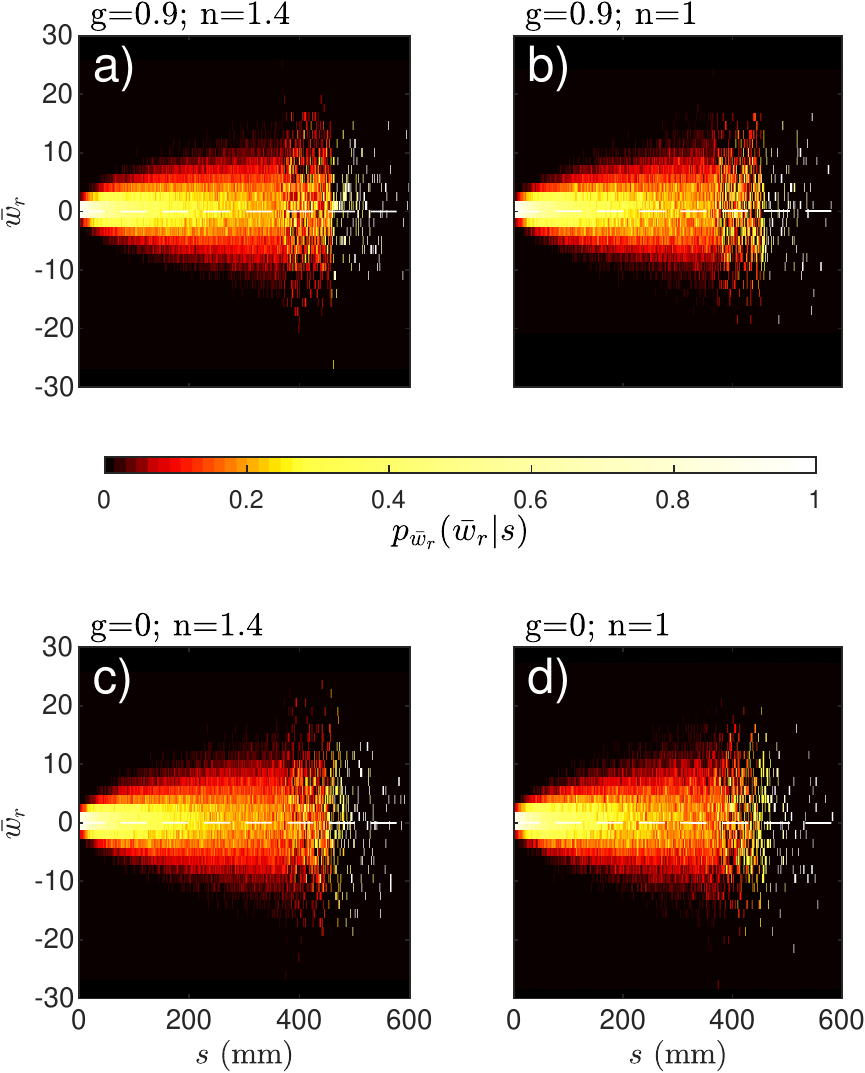}
\caption{Conditional probability density $p_{\bar w_r}(\bar w_r|s)$ 
that a photon's path of length $s$ has mean writhe $\bar w_r$
(no distinction is made in this case between different $\rho$ values).
The horizontal (vertical) axis range in the different panels is the same.}
\label{fig:wrs}
\end{figure}

In Fig. \ref{fig:wrs}
is reported the conditional pdf, $p_{\bar w_r}(\bar w_r|s)$, that a given photon's path of length $s$ has 
a mean writhe $w_r$. 
Here again, the data in Fig. \ref{fig:wrs} are the same as the data in Fig. \ref{fig:wr},
with the distinction between the different~$\rho$ values replaced by the distinction between different~$s$ values.
From the figure panels it appears that the probability to perform a high net number of right or left turns 
increases with~$s$.
This behavior does not depend on the optical parameters $g$ and $n$.

The total computation time utilized to obtain, from the MC data, all $\bar c$ and $\bar w_r$ values 
presented in this contribution was $\sim$50 days.

\subsection{Minimal crossing number and HOMFLY polynomial}
\label{subsec:RHOMFLY}

\begin{figure}[htbp]
\includegraphics[width=\linewidth]{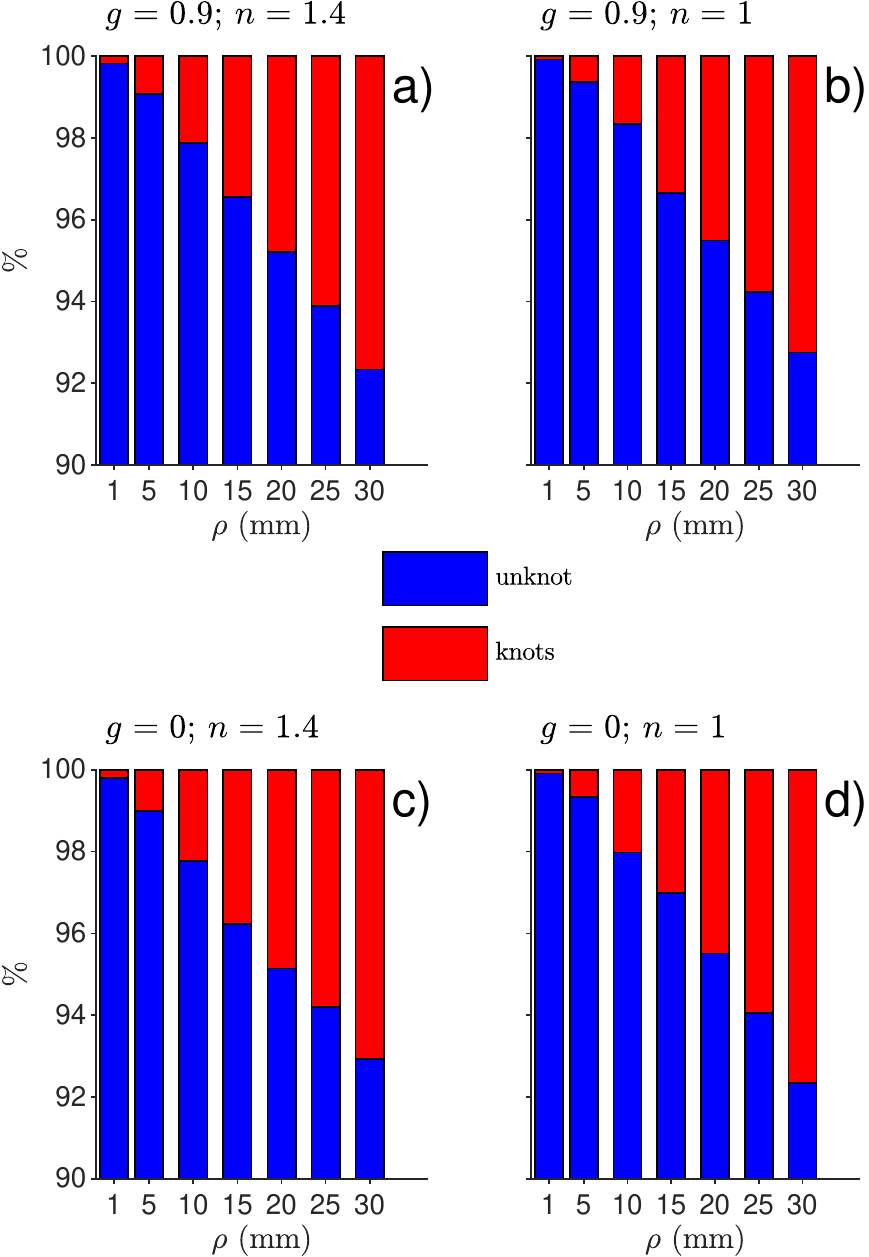}
\caption{Staked bar chart representing the percentage of unknots (blue) and 
non-trivial knots (red) found in each MC simulation.}
\label{fig:KnotsUnknots}
\end{figure}

In Fig. \ref{fig:KnotsUnknots}
are shown the percentages of (unknots and) non-trivial knots for each MC simulation.
These percentages were obtained by evaluating the HOMFLY polynomials different from 1, the value of this invariant for the unknot.
On this figure, it clearly appears that the number of non-trivial knots 
increases with the source/detector spacing $\rho$.

\begin{figure}[htbp]
\includegraphics[width=\linewidth]{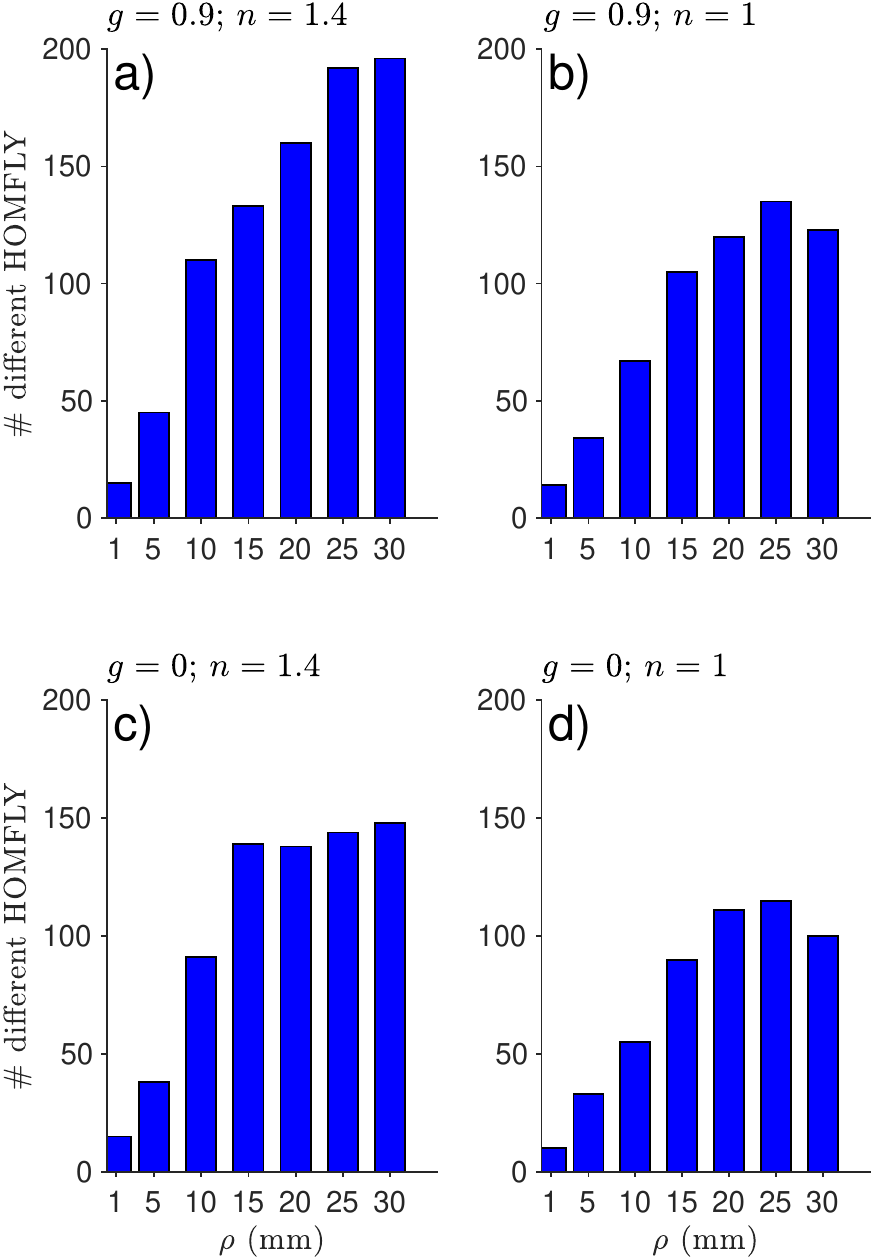}
\caption{Number of different HOMFLY polynomials generated at each MC simulation.}
\label{fig:NumberOfPathFig}
\end{figure}

In Fig. \ref{fig:NumberOfPathFig}
are shown the number of different HOMFLY polynomials 
appearing in each MC simulation. As explained in Sec.~\ref{subsec:HOMFLY}, this number is a very good lower bound on the number of different knot types appearing.
From these data it appears that the topological diversity of the photons' paths is very substantial, and increases with
the interoptode distance $\rho$ (see in particular Fig. \ref{fig:NumberOfPathFig}a, where almost 200 topologically distinct knots are observed for~$\rho=30$). 
Note that the optical parameters $g$ and $n$ do have an influence on the topological complexity of the paths.
This is due to the following fact: when there is no refractive index mismatch, for a given $\rho$ value, less long photons' paths are generated \cite{ref:Martelli2010} and 
thus, less complex paths are observed.
This phenomenon also appears for $g=0$.
The influence of $g$ on the number of different HOMFLY polynomials is due to the fact that large $g$ values implies a higher number of scattering
events, and thus an increased complexity of the paths.

\begin{figure}[htbp]
\includegraphics[width=\linewidth]{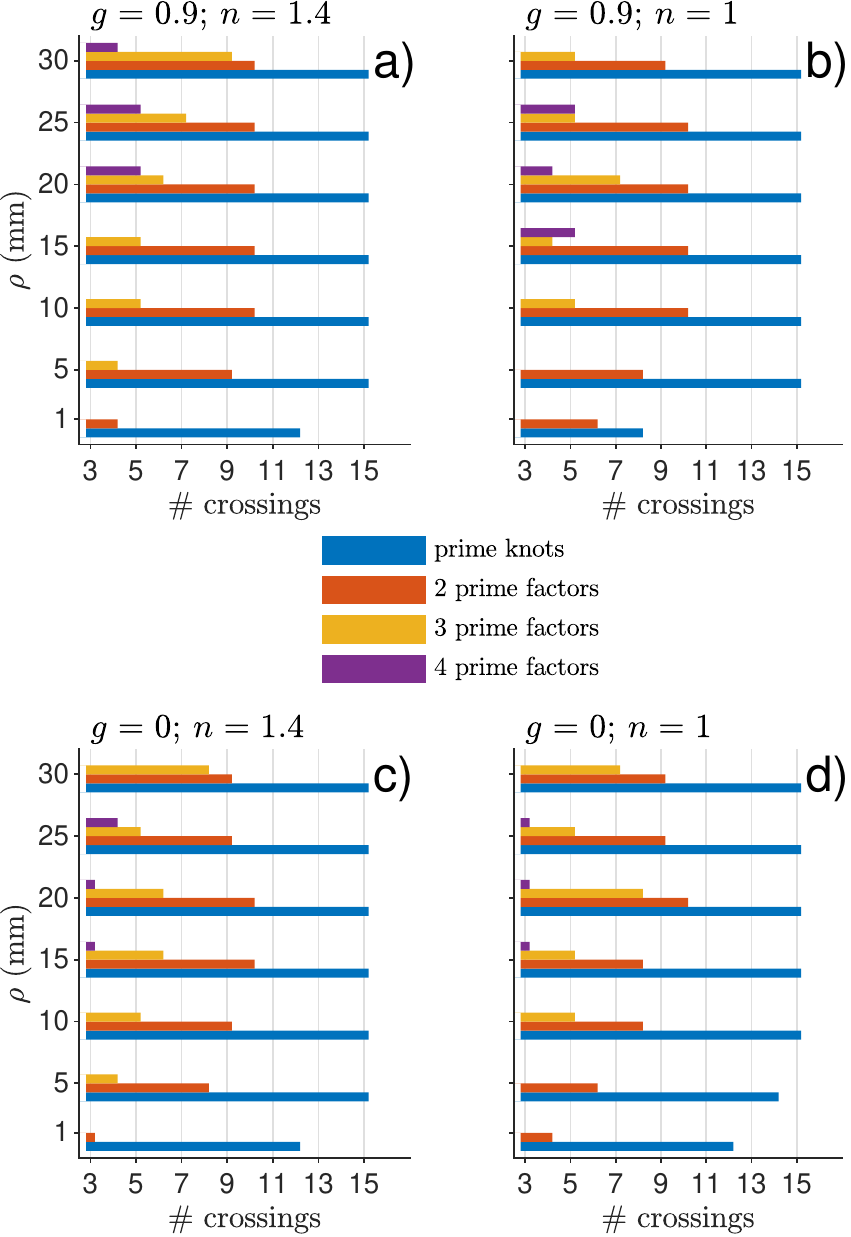}
\caption{Ranges defined by the lowest and highest minimal crossing numbers of the prime factors of the identified knot types,
reported for different $\rho$ values.
}
\label{fig:KnotsMinMaxCrossFig}
\end{figure}

In Fig. \ref{fig:KnotsMinMaxCrossFig}
are shown the ranges defined by the lowest and highest minimal crossing numbers of the prime factors of the generated knot types, sorted according to~$\rho$ values and number of factors. For example, the signification of the top purple bar in Fig. \ref{fig:KnotsMinMaxCrossFig}a is the following: for~$\rho=30$, the knot types with 4 prime factors have factors consisting of 3 and 4-crossing knots (i.e. trefoil and figure eight knots).
For prime knots, all these ranges start at 3 crossings (which is the minimal number of crossings for a non-trivial knot), and reach a maximum of 15 (which is the maximum number of crossings in our tables) with the exception of very short $\rho$ values.
This is due to the fact that at short $\rho$, the paths are also short and they cannot form very intricate knots.
The unknown knot types are not reported in this case because, {\it a fortiori}, their crossing number is unknown.

\begin{figure}[t]
\includegraphics[width=\linewidth]{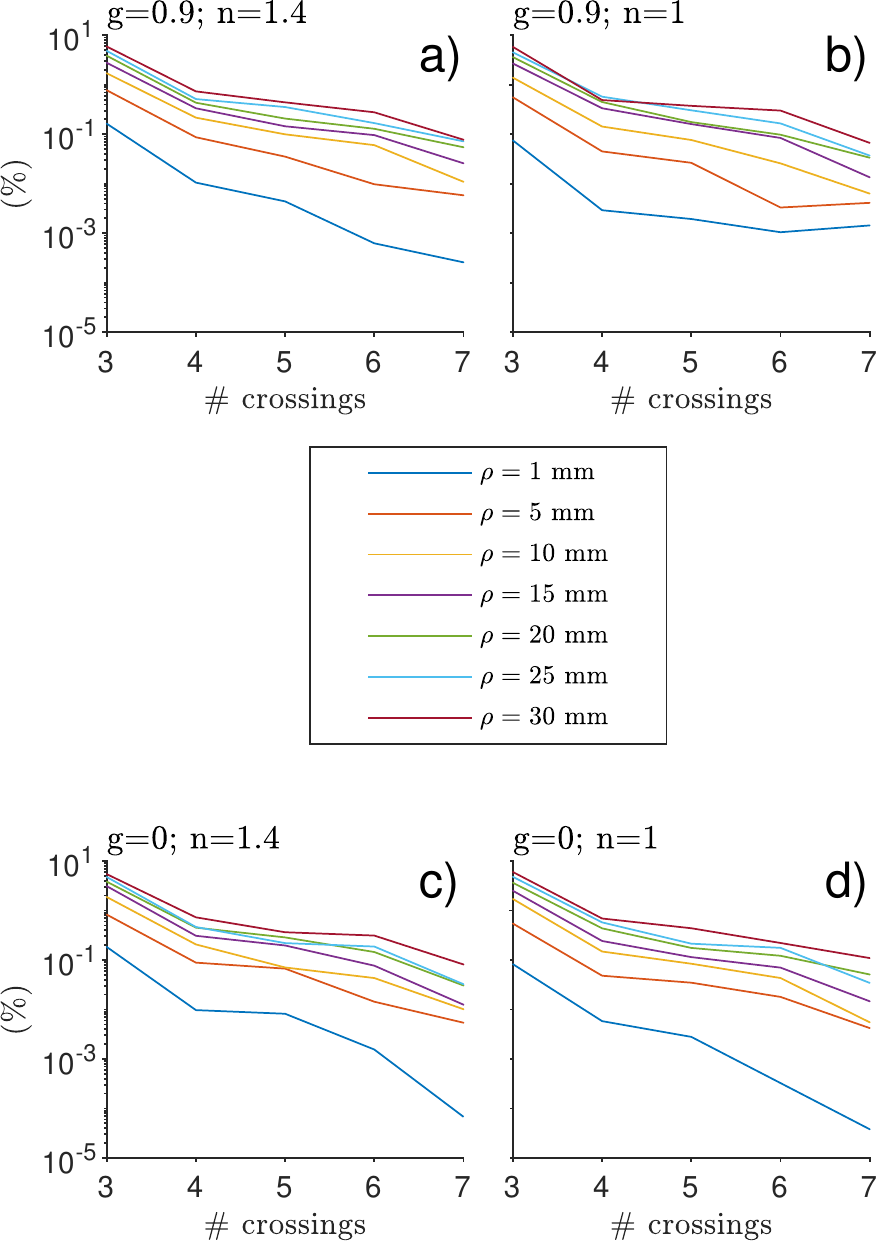}
\caption{Percentages of knots with crossing numbers~$3$ to~$7$.}
\label{fig:SumNumberOfCrossingFig}
\end{figure}

In Fig. \ref{fig:SumNumberOfCrossingFig}
are shown the percentages, among all identified knots, of the knot types with crossing number equal to~$3,4,5,6$ and~$7$. (As demonstrated in Fig.~\ref{fig:KnotsMinMaxCrossFig}, knots with crossing numbers as high as~$15$ were generated, but in such few numbers that the corresponding statistics are not very significant; this is why we restrict ourselves to these values.) Unsurprisingly, knot types with higher crossing number, even though more numerous, are much less likely to appear then simpler knot types such as the trefoil knot. Also, for any fixed crossing number, 
the corresponding knot types are more likely to be generated for higher $\rho$ values. Note that we observe once again the influence of the optical parameters~$g$ and~$n$ already mentioned above.

Finally, the computation of the breadth of the Jones polynomial gave interesting results in terms of high crossing numbers of individual knots. For example, some photons' paths form knots with breadth equal to~$17$, and therefore, at least as many crossings, even for~$\rho$ as short as~15 mm. However, the actual statistics of these breadth are not very informative, and for several reasons. First of all, this bound is sharp for alternating knots, but quite rough for non-alternating knots. And if such knots are rather rare for low crossing number, they quickly outnumber alternating knots when the crossing number becomes large: to give an idea, only~$111\,528$ out of the~$313\,230$ prime knots with at most~$15$ crossings are alternating. Furthermore, as explained above, knots with high crossing numbers are generated in such few numbers that the corresponding graphics are not very telling. For these reason, we did not include a figure illustrating these statistics.

The computation time necessary to obtain the HOMFLY polynomials for all the photons' paths generated in the present contribution was $\sim$246 days.
\section{Discussion and Conclusions}
\label{sec:Discussion}

In the present contribution we defined three parameters to 
describe quantitatively some important geometrical and topological characteristics of photons' paths in random media. 
To this aim, the optical parameters ($\mu_{\rm s}'$ and $\mu_{\rm a}$) for a ``typical'' biological tissue were chosen.
Of course it would be interesting to investigate $\bar c$, $\bar w_r$ and the minimal crossing number for a larger 
set of $\mu_{\rm s}'$ and $\mu_{\rm a}$ values, but,
the prohibitive computation times (see Sec. \ref{sec:Results}) compelled us to study only a representative case.
This means that, in the future, it might be worth putting a greater effort in the improvement of the different algorithms
(e.g. new formulas or high performance computing).
However, this preliminary study already shows us some important characteristics of photons' paths.
In fact, the increasing $\bar c$ values as a function of $s$ (Fig. \ref{fig:acn} and \ref{fig:acns}) and the related increase of the $\bar w_r$ range
(Fig. \ref{fig:wr} and \ref{fig:wrs}), tell us that limited penetration of photon in biological tissues 
in not only determined by absorption \cite{ref:Martelli2016} but might also be influenced by the geometrical complexity of the photons' paths.
This behavior is more important for large $\rho$ (e.g. 30 mm), a situation typically encountered for 
non-invasive studies in humans.
Moreover, the growth of the paths' complexity with increasing $\rho$ also manifests itself through the formation of non-trivial knots.
It remains to be studied what may be the consequences of this phenomenon on optical techniques such e.g. imaging techniques.
This will certainly be interesting matter for future studies.

The global behavior of the results presented here appears to remain stable also for drastic changes in $g$ and $n$.
This is an important point, which shows that we are dealing with a general phenomenon.

In this work, a tissue was studied where photons propagation is governed by
Lambert-Beer's law (typical MCML approach, see Sec. \ref{sec:Methods}.\ref{subsec:PhotonsPaths}).
In real world  however, there are biological tissues (e.g. lung, bone) where photon transport might display an ``anomalous'' behavior \cite{ref:Binzoni2018}.
In such case, photons may sometimes move with extremely large steps but, more often, 
also with a high number of extremely short steps.
For this reason, ``anomalous'' photon transport might generate more complex paths 
(in number and complexity) than the ones presented here,
rendering more difficult the interpretation of optical signals coming from these important human tissues.

On the other side, it remains to provide a systematic study of what may be the 
consequences of complex path formation on optical techniques such e.g. imaging 
techniques. These preliminary results offer an interesting hint 
on these consequences. 
Indeed, the growth of the paths’ complexity with 
increasing $\rho$ suggests an attractive direction of research for imaging 
applications where the use of short or null source-detector distance 
seems to be preferable. 
Imaging techniques based on null or 
short $\rho$ have been proposed based on the advantage of having 
better spatial resolution and contrast \cite{ref:Torricelli2005,ref:Pifferi2008,ref:Mora2015}. 
The present results seem to suggest that the reasons to choose null or 
short $\rho$ may be even more general and intrinsically related to 
geometrical characteristics of photons’ trajectories. 
This will certainly be interesting matter for future studies.

In the present article, we focused on the topological complexity of the photons' paths, as measured by their minimal crossing number. However, it should be pointed out that our methods allow us to do more, and actually identify the corresponding knot type. Using our data, it is therefore possible to study the statistics of the induced knot types, and their dependence on the various parameters. Many such models of random knots have already been studied (see e.g ~\cite{Even-Zohar2017} and references therein), but the mathematical study of knot formation in the presence of a “Lambert-Beer’s” propagation mode and a Henyhey-Greenstein scattering function has, to the best of our knowledge, never been undertaken. Thus, the present study might also lead to interesting new results from a mathematical point of view.

In conclusion, the present contribution clearly shows that a careful study of photons' paths leads 
us far beyond the historical ``banana shape'' interpretation, 
and that a deeper analysis of this topic might improve the interpretation of
optical signals in different domains.


\end{document}